\documentclass[ aip, sd, amsmath,amssymb, preprint,
]{revtex4-1}

\usepackage{graphicx}
\usepackage{dcolumn}
\usepackage{bm}
\usepackage{gensymb}

\usepackage{helvet}

\begin{document}


\title{The Shockley-Queisser limit for nanostructured solar cells}

\author{Yunlu Xu}
\affiliation{Department of Electrical and Computer Engineering, University of Maryland, College Park, MD 20740, USA}
\affiliation{Institute for Research in Electronics and Applied Physics, University of Maryland, College Park, MD 20740, USA}
 
\author{Tao Gong}
\affiliation{Department of Electrical and Computer Engineering, University of Maryland, College Park, MD 20740, USA}
\affiliation{Institute for Research in Electronics and Applied Physics, University of Maryland, College Park, MD 20740, USA}

\author{Jeremy N. Munday}
\email{jnmunday@umd.edu }
\affiliation{Department of Electrical and Computer Engineering, University of Maryland, College Park, MD 20740, USA}
\affiliation{Institute for Research in Electronics and Applied Physics, University of Maryland, College Park, MD 20740, USA}

\date{\today}

\begin{abstract}
\textbf{The Shockley-Queisser limit describes the maximum solar energy conversion efficiency achievable for a particular material and is the standard by which new photovoltaic technologies are compared. This limit is based on the principle of detailed balance, which equates the photon flux into a device to the particle flux (photons or electrons) out of that device. Nanostructured solar cells represent a new class of photovoltaic devices, and questions have been raised about whether or not they can exceed the Shockley-Queisser limit. Here we show that single-junction nanostructured solar cells have a theoretical maximum efficiency of ~42\% under AM 1.5 solar illumination. While this exceeds the efficiency of a non-concentrating planar device, it does not exceed the Shockley-Queisser limit for a planar device with optical concentration. We conclude that nanostructured solar cells offer an important route towards higher efficiency photovoltaic devices through a built-in optical concentration.}
\end{abstract}

\maketitle


In 1961, Shockley and Queisser developed a theoretical framework for determining the limiting efficiency of a single junction solar cell based on the principle of detailed balance equating the incoming and outgoing fluxes of photons for a device at open circuit conditions.\cite{Shockley1961} This model incorporates various light management and trapping techniques including photon recycling, optical concentration, and emission angle restriction.\cite{Shockley1961, Marti1997, Munday2012} It was recently suggested that a nanowire solar cell could exceed the Shockley-Queisser (SQ) limit based on its geometry; \cite{Krogstrup2013} however, without exploiting 3rd generation PV concepts which break the assumptions of Shockley and Queisser (e.g. multi-exciton generation, hot carrier collection, etc), \cite{Green2001} even nanowire solar cells should be bounded by the SQ limit. Here we show that for nanostructured solar cells, the limiting efficiency is identical to that of a planar solar cell with concentrating optics and that the improvement results strictly from an increase in the open circuit voltage. This formalism leads to a maximum efficiency of ~42\% for a nanostructured semiconductor with a bandgap energy of ~1.43 eV (e.g. GaAs) under AM 1.5G illumination.\cite{PV_Handbook}


The SQ limit is reached by applying the principle of detailed balance to the particle flux into and out of the semiconductor. \cite{Shockley1961} For every above bandgap photon that is absorbed by the semiconductor, one electron-hole pair is generated. The maximum possible efficiency is achieved when non-radiative recombination is absent, and all generated carriers are either collected as current in the leads or recombine, emitting a single photon per electron-hole pair. The total generated current is:

\begin{eqnarray}\label{1}
I_{total}=q\left[N_{abs}-N_{emit}\left(V\right)\right]
\end{eqnarray}
where q is the charge of an electron, and $N_{abs}$ and $N_{emit}$ are the numbers of photons per unit time that are absorbed or emitted by the photovoltaic device, respectively. These rates can be calculated as: \cite{Marti1997}

\begin{eqnarray}\label{2}
\begin{aligned}
N\left(\theta_{max},V,T\right)=&\int_{0}^{\infty}\int_{\phi=0}^{2\pi}\int_{\theta=0}^{\theta_{max}}\sigma_{abs}\left(\theta,\phi,E\right)
\times F\left(E,T,V\right)\cos\left(\theta\right)\sin\left(\theta\right)d\phi d\theta dE\
\end{aligned}
\end{eqnarray}
where $\sigma_{abs}(\theta,\phi,E)$ is the absorption cross-section, $F(E,T,V)$ is the spectral photon flux, and $\theta_{max}$ is the maximum angle for absorption (for $N_{abs}$) or emission (for $N_{emit}$). For a bulk planar cell, the absorption cross-section is given by $\sigma_{abs}(\theta,\phi,E)=A_{cell}\times a(\theta,\phi,E)$, where $A_{cell}$ is the top illuminated surface area of the cell and $a(\theta,\phi,E)$ is the angle dependent probability of photon absorption for incident photons of energy $E$. In the simplest case, $a(\theta,\phi,E)$ is a step-function going from 0 for $E<E_g$ to 1 for $E\geq E_g$. The spectral photon flux can be obtained from the generalized Planck blackbody law:\cite{Wurfel1995}

\begin{eqnarray}\label{3}
F\left(E,T,V\right)=\frac{2n^2}{h^3 c^2}\frac{E^2}{e^{\frac{E-qV}{k_bT}}-1}
\end{eqnarray}
where $h$ is Planck's constant, $k_b$ is Boltzmann's constant, $c$ is the speed of light, $n$ is the refractive index of the surroundings, which is usually taken to be vacuum ($n=1$), and $qV$ characterizes the quasi-Fermi level splitting when describing emission from the cell. The incoming flux from the sun can be obtained from experimental data (e.g. AM 1.5 solar spectrum \cite{PV_Handbook}) or from the blackbody expression above with $V=0$ and where $\theta_{max}=\theta_s=0.267\degree$ is the acceptance half-angle for incident light from the sun at temperature $T=T_s=5760 K$. The outgoing flux from the cell is given by Eq. [\ref{2}] for a cell temperature $T_c=300 K$, operating voltage $V$, and emission half-angle $\theta_{max}=\theta_c=90\degree$. At open circuit conditions, there is no current extracted, and the current balance equation becomes

\begin{eqnarray}\label{4}
\begin{aligned}
0=&qN\left(\theta_s,T_s,V=0\right)+qN\left(\theta_c,T_c,V=0\right)
-qN\left(\theta_c,T_c,V=V_{oc}\right)
\end{aligned}
\end{eqnarray}
where the middle term corresponds to absorption due to emission from the ambient surroundings, also at $T=300 K$; however, this term is much smaller than the flux from the sun. Thus, the light generated current is given by $I_L=qN(\theta_s,T_s,V=0)$ and the dark current, in the radiative limit, is given by $I_0=I_R \exp(\frac{qV}{k_B T_c})=qN(\theta_c,T_c,V)$, where $I_R$ is the reverse saturation current. Solving the above expression for the voltage yields the common expression for the open circuit voltage:\cite{Shockley1961, PV_Handbook}

\begin{eqnarray}\label{5}
\begin{aligned}
V_{oc}=\frac{k_BT_c}{q}\ln\left(\frac{I_L}{I_R}+1\right)\approx\frac{k_BT_c}{q}\ln\left(\frac{I_L}{I_R}\right)
\end{aligned}
\end{eqnarray}
which is valid for both bulk planar solar cells and nanostructured solar cells with the appropriate absorption cross-sections as described in the next section. 

\subsection*{Results}
\noindent\textbf{Nanostructured solar cells with built-in optical concentration.} To achieve the maximum efficiency, we need to increase the light generated current compared to its bulk form or reduce the reverse saturation current to increase $V_{oc}$. For any absorbing structure, Eqs.[\ref{2},\ref{3},\ref{4},\ref{5}] can be used to determine the resulting $V_{oc}$ numerically; however, for the limiting case, we will consider a simple analytical expression. For maximum $V_{oc}$, we want the absorption cross-section to be maximized for angles near normal incidence $0\le\theta\le\theta_m$ and minimized for all other angles $\theta_m\le\theta\le\theta_c$, where $\theta_s\le\theta\le\theta_c$ and $\theta_m$ is some angle defined by the structure. We can define this piece-wise function for the absorption cross-section as: $\sigma_{abs}(\theta:0\to\theta_m)=\sigma_{max}$ and $\sigma_{abs}(\theta:\theta_m\to\theta_c)=\sigma_{min}$, which allows us to perform the solid angle integration to determine the light and dark currents:

\begin{eqnarray}\label{6}
\begin{aligned}
I_L=&qN\left(\theta_s,T_s,V=0\right)\\
=&\sigma_{max}\int_{E_g}^{\infty}\int_{\phi=0}^{2\pi}\int_{\theta=0}^{\theta_{s}}F\left(E,T_s,V=0\right)
\times \cos\left(\theta\right)\sin\left(\theta\right)d\phi d\theta dE\
\\=&
\frac{\sigma_{max}}{A_{cell}}I_{L,0}
\end{aligned}
\end{eqnarray}
where $\sigma_{abs}=0$ for $E<E_g$, $I_{L,0}$ is the light generated current for a bulk cell of area ${A_{cell}}$, and

\begin{eqnarray}\label{7}
\begin{aligned}
I_R=&qN\left(\theta_c,T_c,V=0\right)\\
=&\frac{\pi q\sigma_{min}}{2}\left[\cos\left(2\theta_m\right)-\cos\left(2\theta_c\right)\right]\int_{E_g}^{\infty}F\left(E,T_c,V=0\right)dE
\\&+\frac{\pi q\sigma_{max}}{2}\left[1-\cos\left(2\theta_m\right)\right]\int_{E_g}^{\infty}F\left(E,T_c,V=0\right)dE
\\=&\frac{\sigma_{max}+\sigma_{min}+\left(\sigma_{min}-\sigma_{max}\right)\times\cos\left(2\theta_m\right)}{2A_{cell}}I_{R,0}
\end{aligned}
\end{eqnarray}
where $I_{R,0}$ is the reverse saturation current for a bulk cell. Substituting these expressions into  Eq. [\ref{5}], we have 
\begin{eqnarray}\label{8}
\begin{aligned}
V_{oc}\approx&\frac{k_BT_c}{q}\ln\left[\frac{2\sigma_{max}}{\sigma_{max}+\sigma_{min}+(\sigma_{min}-\sigma_{max})\cos(2\theta_m)}\right]
+\frac{k_BT_c}{q}\ln\left[\frac{I_{L,0}}{I_{R,0}}\right]
\\=&\frac{k_BT_c}{q}\left[\ln\left(\frac{I_{L,0}}{I_{R,0}}\right)+\ln\left(X\right)\right]
\end{aligned}
\end{eqnarray}

where 

\begin{eqnarray}\label{9}
X=\frac{2\sigma_{max}}{\sigma_{max}+\sigma_{min}+(\sigma_{min}-\sigma_{max})\cos(2\theta_m)}
\end{eqnarray}

Thus, the open circuit voltage for a nanostructured device takes on the same form as the open circuit voltage for a macroscopic concentrating system, where X is the concentration factor.\cite{PV_Handbook} For maximum concentration, we consider the limit as $\theta_m\to\theta_s$ and $\sigma_{min}\to0$, yielding

\begin{eqnarray}\label{10}
X=\frac{2}{1-\cos(2\theta_s)}\approx46,050
\end{eqnarray}
which is the same as the maximum concentration factor that is obtained for a macroscale concentrator and results in a maximum solar energy conversion efficiency of $\sim42\%$. For practical devices it is reasonable to assume a minimum value of $\sigma_{min}$  corresponding to the geometric cross-section of the device, $\sigma_{min}\to\sigma_{geo}$. For this case, and with $\cos(2\theta_m)=\cos(2\theta_s)\approx1$, we get $X=\sigma_{max}/\sigma_{geo}$, and the open circuit voltage reduces to:
 
\begin{eqnarray}\label{11}
V_{oc}=\frac{k_BT_c}{q}\ln\left[\frac{\sigma_{max}}{\sigma_{geo}}\left(\frac{I_{L,0}}{I_{R,0}}\right)\right]
\end{eqnarray}

Finally, the power conversion efficiency is given by $\eta=I_LV_{oc}FF/P_{in}$, where $FF$ is the fill-factor, which can be obtained from the $I-V$ characteristic defined by Eq. [1], and $P_{in}$, is the incident power from the sun. We note that the area used to calculate $P_{in}$ is determined by the illumination area and not the geometric cross-section, which would lead to under counting the number of incident photons.
In general, optical concentration can be achieved using lenses, mirrors, or unique optical nanostructures (see Fig. 1(a)). A nanostructured solar cell can result in optical concentration that is similar to the concentration obtained using lens or parabolic mirrors but relies on the wave nature of light. Fig. 1 (b) shows the power conversion efficiency of recently reported vertically aligned nanowire-based PV cells. \cite{Putnam2010, Yang2011, Wang2011, Jung2012, Huang2012, Kendrick2010, Nguyen2011, Krogstrup2013, Mariani2013, Cirlin2010, Nakai2013, Mariani2011, Wallentin2013, Cui2013, Yoshimura2013, Goto2009} The optical and geometrical cross-sections are extracted from the current density data and from the geometrical information provided within the references. The vast majority of the experiments are focused on Si, GaAs and InP radial or axial junction nanowire arrays fabricated with various techniques, such as MBE, MOVCD, reactive-ion etching, etc. Generally, $X=\frac{\sigma_{max}}{\sigma_{geo}}$ is found to fall in the range of 1-10 for these structures; however, the actual concentration factor is likely significantly smaller if $\sigma_{min}>\sigma_{geo}$. Additionally, the reduced efficiency in these nanowire structures compared to the theoretical limit is due to significant surface recombination and device and material constraints that could be improved with further experimental development.

\bigskip
\noindent\textbf{The effect of entropic losses on $V_{oc}$.} Next we consider an alternative, but equivalent, approach to understanding the maximum efficiency of a nanostructured PV device by considering the energetic and entropic loss mechanisms. \cite{Hirst2010, Atwater2012, Rau2014} The generalized Planck equation can be used to determine the open circuit voltage of a solar cell operating at the maximum efficiency limit: \cite{Henry1980, Ruppel1980,Hirst2010}

\begin{eqnarray}\label{12}
\begin{aligned}
V_{oc}=&\frac{E_g}{q}\left(1-\frac{T_c}{T_s}\right)+\frac{k_BT_c}{q}\ln\left(\frac{\gamma_s}{\gamma_c}\right)
-\frac{k_BT_c}{q}\ln\left(\frac{\Omega_{emit}}{\Omega_{abs}}\right)
\end{aligned}
\end{eqnarray}
where $\gamma_s$ and $\gamma_c$ are blackbody radiation flux terms that depend on $E_g$, $T_s$, and $T_c$. The first term represents a voltage drop related to the conversion of thermal energy into work (sometimes called the Carnot factor). The second term occurs from the mismatch between Boltzmann distributions at $T_c$ and $T_s$.\cite{Markvart2007} The third term is the voltage loss due to entropy generation as a result of a mismatch between the absorption solid angle and the emission solid angle of the cell. This third term represents a voltage drop of $\sim0.28$ V, which can be recovered if ${\Omega_{emit}}={\Omega_{abs}}$.\cite{Braun2013,Kosten2014}

The most common way to recover the entropy loss due to the mismatch between the absorption and emission solid angles is through optical concentration (Fig. 2(a)). For a planar solar cell without optical concentration, the absorption solid angle corresponds to the sun's angular extent, i.e. $\Omega_{abs}=2\pi\left(1-\cos\left(\theta_s\right)\right)=6.85\times 10^{-5}$ sr. However, emission from the cell occurs over ${\Omega_{emit}}=4\pi$. The addition of a back reflector reduces the emission solid angle to ${\Omega_{emit}}=2\pi$, resulting in a slight voltage improvement. \cite{Marti1997} For more substantial voltage improvements, optical concentration is necessary. Optical concentration enables the absorption solid angle to exceed the sun's solid angle and approach the cell's emission solid angle (Fig. 2(a)), which could largely increase the $V_{oc}$.

Properly designed photovoltaic nanostructures can have the same effect, reducing the entropy generation by either increasing ${\Omega_{abs}}$ or by reducing ${\Omega_{emit}}$ in an attempt to achieve ${\Omega_{emit}}={\Omega_{abs}}$ (Fig. 2(b)). From a device point-of-view, ${\Omega_{abs}}$ is related to the light generated current density, $J_L=I_L/A$, and ${\Omega_{emit}}$ is related to the reverse saturation current density, $J_R=I_R/A$. Because the $V_{oc}$ depends on their ratio (see Eq. [5]), increasing ${\Omega_{abs}}$ will have the same affect as decreasing ${\Omega_{emit}}$. Thus, the voltage improvement can equivalently be seen from the thermodynamics of reduced entropy generation or from the device aspects of the pn-junction.

According to Kirchhoff's law, the emissivity and absorptivity of a solar cell are equal in thermal equilibrium. \cite{Araújo1995, Marti1997} For a standard cell without back reflector, the device can absorb the incident power from all directions and hence will emit in all directions (Fig. 3(a)). The addition of a back reflector reduces both absorption and emission from the back surface (Fig. 3(b)); however, this has no effect on the absorption of the incident solar power because no illumination is coming from the back. Thus, $I_L$ is unaffected by the addition of the back reflector but $I_R$ is reduced.\cite{Note1} An ideal nanostructure would allow for absorption only over the range of angles corresponding to the incident illumination of the source, i.e. the sun (Fig. 3(c)). The current-voltage characteristics for these devices show that a back reflector yields a $\sim2\%$ increase in efficiency over the traditional planar device, and an ideal nanostructure yields a $\sim11\%$ improvement, resulting in a $\sim42\%$ efficient device.

\bigskip
\noindent\textbf{Numerical simulation of nanowire PV.} We simulated a bulk (80 $\mu m$ thick) GaAs solar cell and a nanowire solar cell with the same thickness using S4\cite{Liu2012} to solve the detailed balance expression numerically.\cite{Sandhu2013,Sandhu2014} For simplicity, we used the blackbody spectrum in the following calculations. The nanowires are embedded within an index of 2.66 and both the nanowire and planar structures are coated with a double-layer antireflection coating (52 nm of n=2.66 and 98 nm of 1.46). The antireflection coating is designed to maximize the efficiency of the bulk GaAs cell. The integrated short circuit current density is almost identical for both cases ($<1\%$ difference); however, the emitted power density is significantly different. Because a large amount of the radiated power is near the bandgap, the lower absorption rate near the bandgap that occurs with the nanowire structure leads to a decrease in emission. This effect is demonstrated in Fig. 4(d), where the bulk cell has a higher reverse saturation current density compared to the nanowire cell with same thickness. The reverse saturation current of the nanowire cell decreases by 3.46\%, and the absorption increases by 0.38\%. As a result, the $V_{oc}$ increases by 10 mV due to these combined effects in the nanowire device, and thus, the nanowire solar cell has a slightly higher efficiency than the bulk device (28.22\% vs. 28.09\%).

Ideally, an optical structure should be designed to minimize absorption for angles greater than $\theta_s$, particularly near the semiconductor bandgap, which is where the emission is peaked. To emphasize this effect, we consider a smaller radius nanowire (40 nm), which will have increased optical concentration. In order to minimize the loss in photogenerated current, the periodicity is decreased to 200 nm, and the nanowire length is set to 2 $\mu$m, which is a reasonable thickness for a GaAs cell. Fig. 4(c) shows this device whose absorption near the bandgap is limited so that the reverse saturation current density is one order of magnitude smaller than that of the bulk cell (Fig. 4(d)). This nanostructuring leads to the reverse saturation current decreasing from $8.751\times10^{-18}$ to  $9.946\times10^{-19}$ A/m$^2$. Although the absorption is also decreased ($J_L$ decreased from $362.68$ to $237.55$ A/m$^2$), the $V_{oc}$ is increased from $1.169$ V to $1.214$ V, showing an improvement of 45 mV in $V_{oc}$. This result suggests that nanostructures that incorporate more complexity may be needed to yield higher $V_{oc}$'s without loss in $I_L$. 

\subsection*{Discussion}
While the overall performance of nanostructed solar cells is still bounded by the SQ limit, one must consider the built-in optical concentration when applying this theory. Recently an InP nanowire solar cell was found to have a $V_{oc}$ in excess of the record InP planar device. \cite{Wallentin2013, Green2014} This improvement is likely the result of the built-in optical concentration, which leads to higher carrier densities and hence a higher $V_{oc}$. Although the best devices to date are $<14\%$ efficient, \cite{Putnam2010, Yang2011, Wang2011, Jung2012, Huang2012, Kendrick2010, Nguyen2011, Krogstrup2013, Mariani2013, Cirlin2010, Nakai2013, Mariani2011, Wallentin2013, Cui2013, Yoshimura2013, Goto2009} there is great potential for improvement, which could allow nanowire solar cells to exceed $40\%$ solar power efficiency. Here we have shown that besides the possibility of improved carrier collection that has been previously reported,\cite{Kayes2005, Colombo2009, Kelzenberg2010} another key advantage of nanostructured solar cells over planar ones is that the optical concentration is already built-in, yielding the possibility of higher efficiencies than planar devices. 

In conclusion, we have used the principle of detailed balance to determine the maximum efficiency for nanostructured photovoltaic devices. The ideal nanostructured devices result in an efficiency of 42\%, which is equivalent to the result of Shockley and Queisser when considering full optical concentration. This improvement comes strictly from an improvement of the open circuit voltage, and not from an improvement in the current. For future nanostructured devices to take advantage of these benefits, high quality surface passivation and reduced non-radiative recombination are needed. From an optical design point-of-view, nanostructures should be created that have limited absorption for angles and wavelengths that do not match the incident illumination. When this condition is achieved, new high efficiency nanostructured PV devices will be possible.




%

\newpage
\begin{figure} [t]	
\includegraphics[width=12cm]{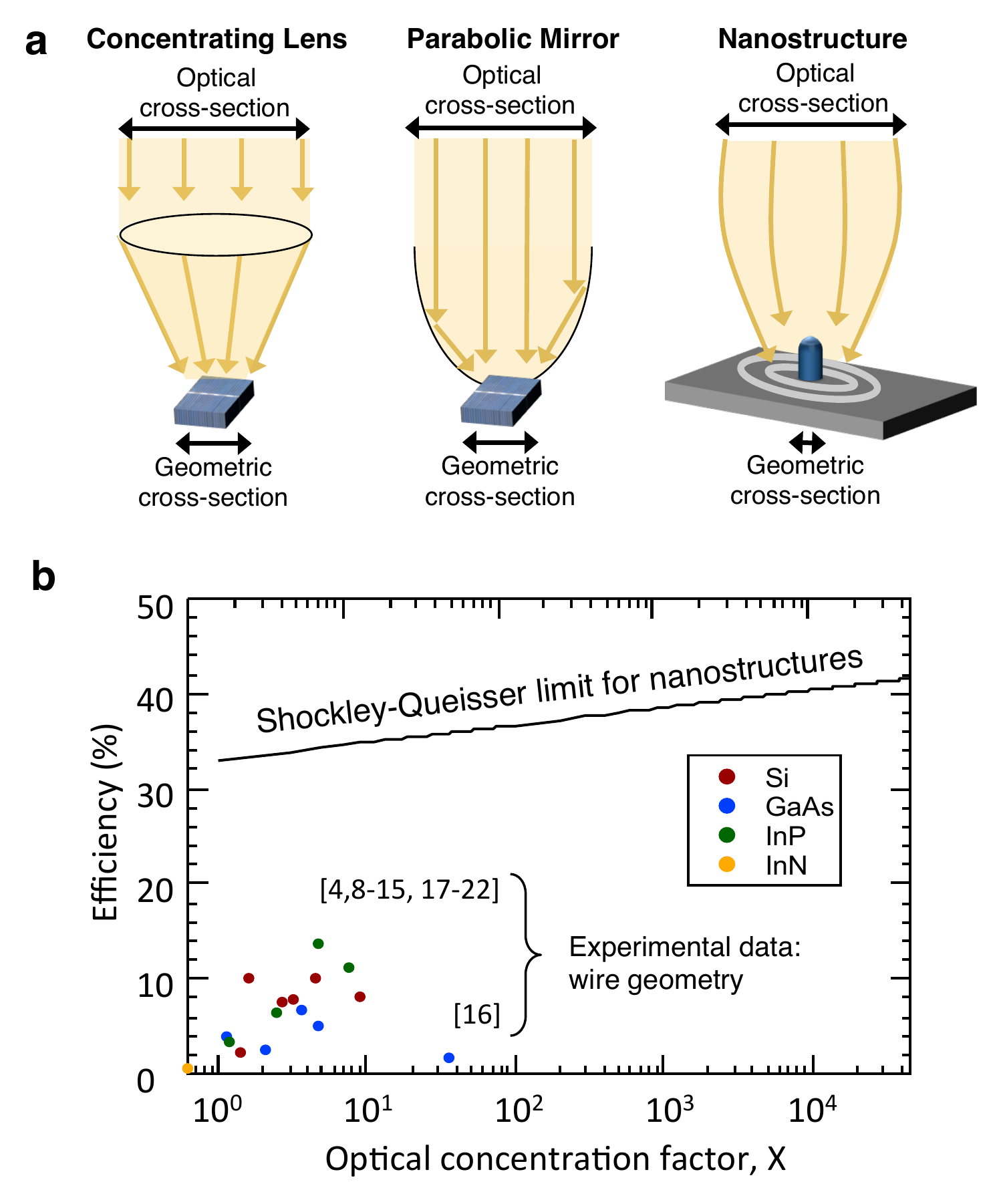}
\end{figure}
\noindent\textbf{Figure 1 $|$ The Shockley-Queisser limit for nanostructures.} (\textbf{a}) Schematic of the optical concentration implemented by a concentrating lens, parabolic mirror, and using a nanostructure itself (self concentration). (\textbf{b}) The efficiencies of cells with optical concentration. The solid line is the theoretical limit of nanostructured PV devices based on detailed balance, whereas individual dots represents experimental data reported in the literature. \cite{Putnam2010, Yang2011, Wang2011, Jung2012, Huang2012, Kendrick2010, Nguyen2011, Krogstrup2013, Mariani2013, Cirlin2010, Nakai2013, Mariani2011, Wallentin2013, Cui2013, Yoshimura2013, Goto2009}

\newpage

\begin{figure} [t]	
\includegraphics[width=12cm]{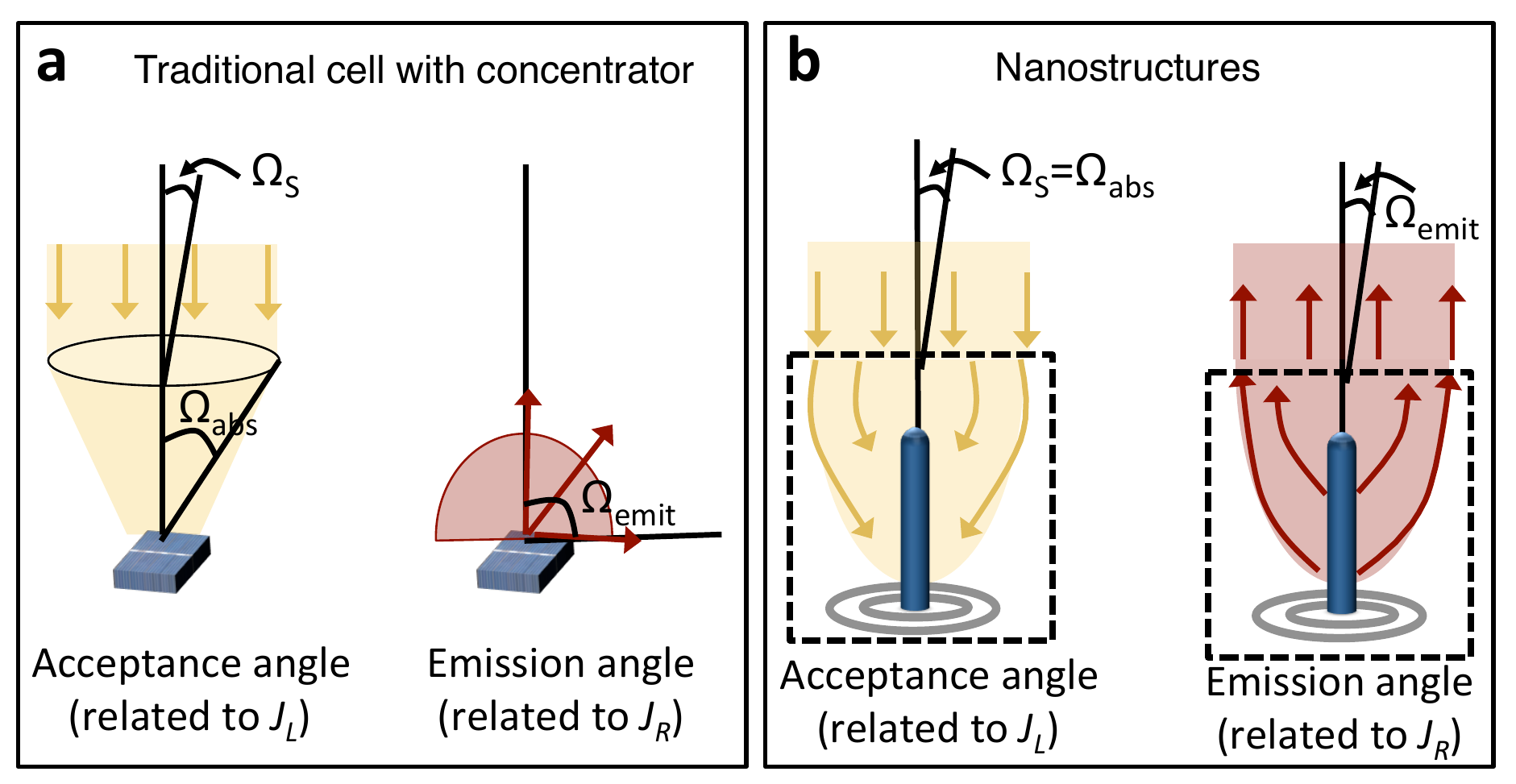}
\end{figure}
\noindent\textbf{Figure 2 $|$ Nanostructures can reduce the mismatch between absorption and emission angles.} (\textbf{a}) A traditional planar solar cell with concentrator increases ${\Omega_{abs}}$ to approach ${\Omega_{emit}}$, thus reducing the entropy generation caused by their mismatch. (\textbf{b}) Similarly, a nanostructured solar cell can reduce the difference between ${\Omega_{abs}}$ and ${\Omega_{emit}}$.

\newpage
\begin{figure} [t] 
\includegraphics[width=12cm]{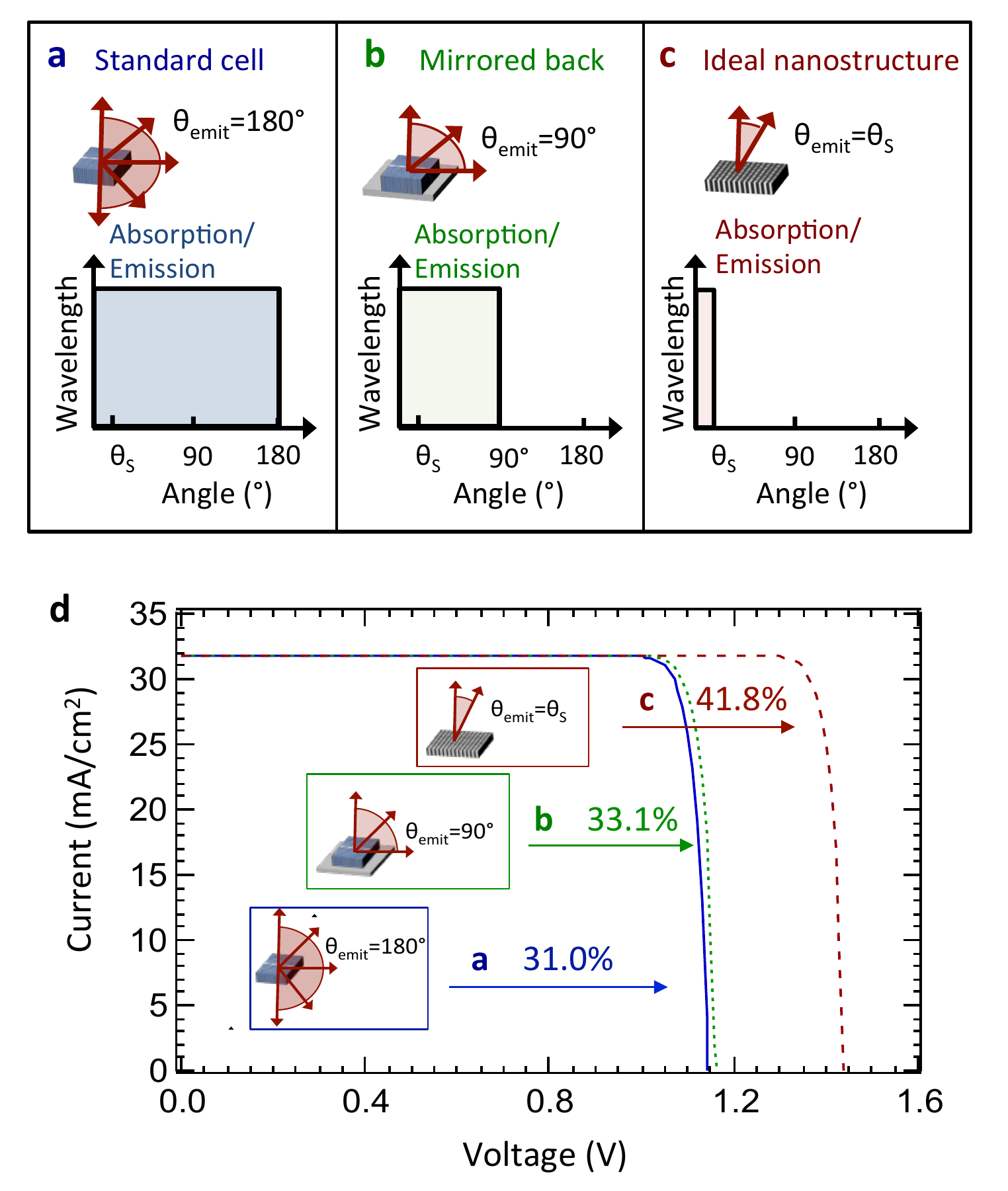}
\end{figure}
\noindent\textbf{Figure 3 $|$ Modification of absorption and emission results in an ideal PV nanostructure achieving $>$40\% power conversion efficiency.} Emission and absorption for (\textbf{a}) slab without back reflector, (\textbf{b}) slab with back reflector and (\textbf{c}) ideal nanostructured cell. The emission and absorption are represented in terms of their half-angle, $\theta$. Absorption/emission over all angles (standard cell) corresponds to $\theta=180\degree$; however, the illumination from the sun is only over a subset of half-angles from 0 to $\theta_s$. Thus, the mismatch between $\theta_s$ and $\theta_{emit}$ results in a decreased voltage. (\textbf{d}) I-V curves corresponding to the three structures (a-c). All structures are illuminated with the AM1.5G spectrum and show increased $V_{oc}$ as $\theta_{emit}\rightarrow \theta_s$.

\newpage
\begin{figure} [t]	
\includegraphics[width=12cm]{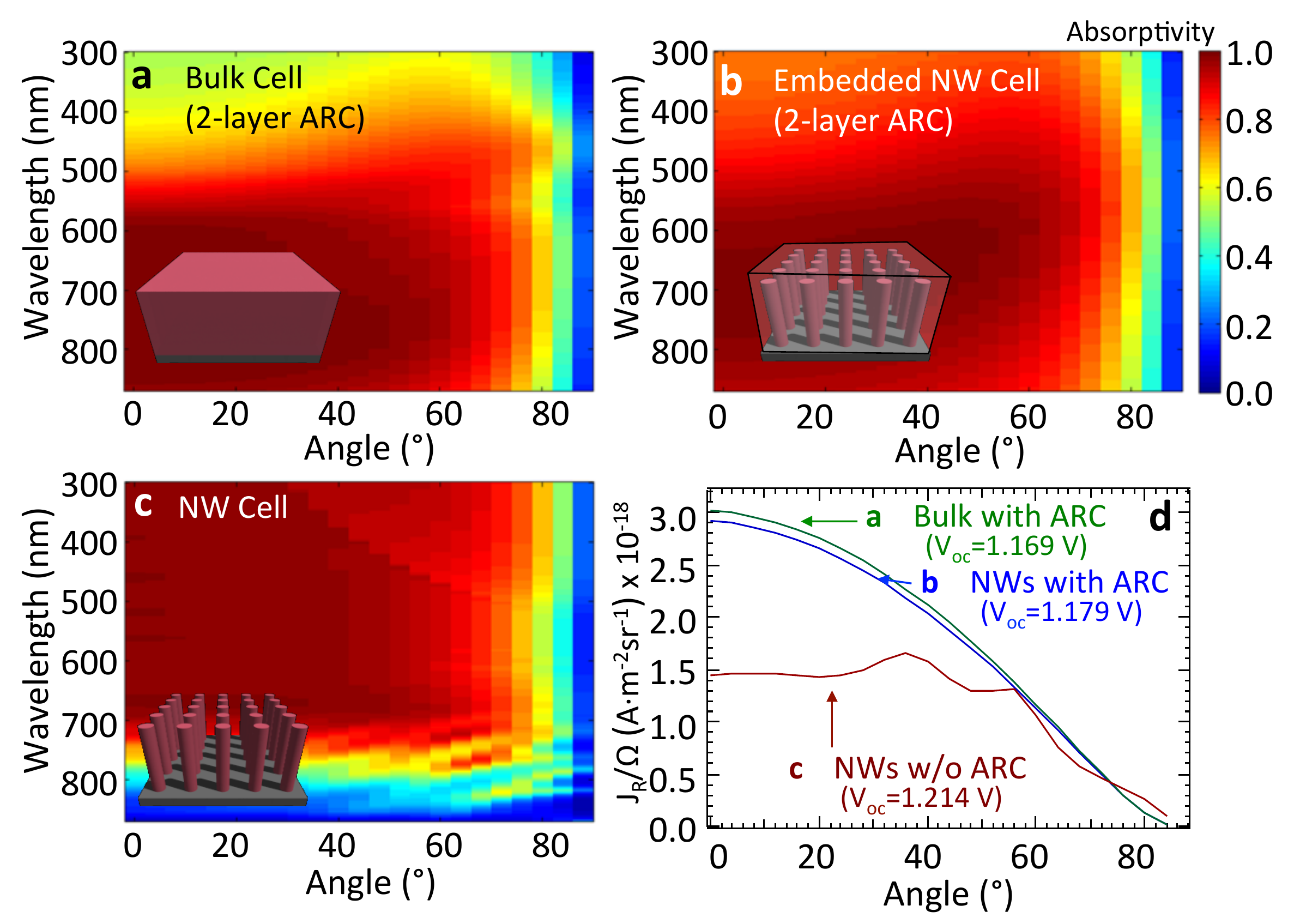}
\end{figure}
\noindent\textbf{Figure 4 $|$ Reduced dark current in nanowire structures.} Angular dependence of the absorption spectrum for (\textbf{a}) a bulk (80 $\mu m$ thick) GaAs solar cell, (\textbf{b}) a GaAs nanowire solar cell (embedded in a dielectric) with a period of 300 nm, a radius of 75 nm, and length of 80 $\mu$m, and (\textbf{c}) a GaAs nanowire solar cell with a period of 200 nm, a radius of 40 nm, and a length of 2 $\mu$m. The devices in (a) and (b) have a double-layer ARC on top, and all cells have a perfect back reflector. The nanowire solar cells have decreased absorption (and hence emission) near the bandedge for angles $>\theta_s$. (\textbf{d}) The current density corresponding to the three structures (a-c) decreases, showing an improved $V_{oc}$ for the nanowire devices.

\end{document}